\documentclass[letterpaper,10pt,conference]{ieeeconf}
\IEEEoverridecommandlockouts
\usepackage[colorlinks=true,linkcolor=blue,citecolor=blue,urlcolor=blue]{hyperref}
\usepackage{cite}
\usepackage{amsmath,amssymb,amsfonts}
\usepackage{algorithmic}
\usepackage{graphicx}
\usepackage{textcomp}
\usepackage{adjustbox}
\usepackage{url}
\usepackage[dvipsnames]{xcolor}
\usepackage[utf8]{inputenc}
\usepackage[normalem]{ulem}
\usepackage{comment}
\usepackage{subcaption}
\usepackage{threeparttable}
\usepackage{soul}
\usepackage{float}
\usepackage{booktabs}
\usepackage[most]{tcolorbox}
\definecolor{lightgray}{gray}{0.95}
\tcbset{
  promptbox/.style={
    colback=lightgray,       
    colframe=black!50,       
    fontupper=\ttfamily\small,
    leftrule=0.5pt,
    rightrule=0.5pt,
    toprule=0.5pt,
    bottomrule=0.5pt,
    boxsep=4pt,
    arc=2pt,
    outer arc=2pt,
  }
}

\def\BibTeX{{\rm B\kern-.05em{\sc i\kern-.025em b}\kern-.08em
    T\kern-.1667em\lower.7ex\hbox{E}\kern-.125emX}}

\title{\LARGE \bf
Sentiment Matters: An Analysis of 200 Human-SAV Interactions}

\author{Lirui Guo$^{1}$, Michael G. Burke$^{2}$, and Wynita M. Griggs$^{1,2}$
\thanks{*This work was supported by an Australian Government Research Training Program (RTP) Scholarship.}
\thanks{$^{1}$Department of Civil and Environmental Engineering, Monash University, Clayton, VIC 3800, Australia.}
\thanks{$^{2}$Department of Electrical and Computer Systems Engineering, Monash University, Clayton, VIC 3800, Australia.}
\thanks{Email: \{Lirui.Guo, Michael.G.Burke, Wynita.Griggs\}@monash.edu}
}

\begin{document}

\maketitle
\thispagestyle{empty}
\pagestyle{empty}

\begin{abstract}
Shared Autonomous Vehicles (SAVs) are likely to become an important part of the transportation system, making effective human–SAV interactions an important area of research. This paper introduces a dataset of 200 human-SAV interactions to further this area of study. We present an open-source human–SAV conversational dataset, comprising both textual data (e.g., 2{,}136 human–SAV exchanges) and empirical data (e.g., post-interaction survey results on a range of psychological factors). The dataset's utility is demonstrated through two benchmark case studies: First, using random forest modeling and chord diagrams, we identify key predictors of SAV acceptance and perceived service quality, highlighting the critical influence of response sentiment polarity (i.e., perceived positivity). Second, we benchmark the performance of an LLM-based sentiment analysis tool against the traditional lexicon-based TextBlob method. Results indicate that even simple zero-shot LLM prompts more closely align with user-reported sentiment, though limitations remain. This study provides novel insights for designing conversational SAV interfaces and establishes a foundation for further exploration into advanced sentiment modeling, adaptive user interactions, and multimodal conversational systems.
\end{abstract}

\section{Introduction}

The rapid advancement of autonomous vehicle (AV) technology is transforming urban mobility, with shared autonomous vehicles (SAVs) anticipated to play a central role in future transportation systems \cite{merfeld_carsharing_2019, dai_impacts_2021}. As SAVs transition from controlled experimental settings to real-world deployment, understanding the dynamics of human interaction with these vehicles and identifying key predictors of user acceptance becomes increasingly critical \cite{zhangRolesInitialTrust2019, guoNewFrameworkPredict2025}. In the context of high autonomy levels (SAE Levels 4/5), where vehicles are operated without the involvement of human drivers,  successful adoption is influenced by a combination of advanced technical capabilities and effective human-centric interactions driven by seamless communication interfaces \cite{ruijtenEnhancingTrustAutonomous2018}. In particular, conversational interactions may serve as a key interface for trip requests, status updates, and in-ride assistance.  

However, current research has largely overlooked the role of conversational dialogue in shaping psychological aspects such as psychological ownership, perceived service quality, and overall user acceptance. Prior studies on human–SAV interactions have primarily utilized structured surveys, driving simulations (e.g., Wizard-of-Oz protocols, motion-based simulators, and virtual reality environments), and assessments focusing on vehicle interior and exterior design \cite{zhangRolesInitialTrust2019, zouRoadVirtualReality2021, zhangToolsPeersImpacts2023, wangExploringImpactConditionally2024}. Although valuable, these methods typically focus on static or predefined scenarios, offering limited insight into the real-time dynamic conversational interactions users will experience in fully autonomous vehicles.

Recent attempts to bridge this gap have begun assembling conversational interaction datasets. For instance, \cite{okur-etal-2020-audio} developed a multimodal Wizard-of-Oz dataset consisting of 30 hours of in-cabin interactions (10{,}590 utterances) collected from 30 participants during a scavenger-hunt activity, integrating speech, audio, and visual data to enhance intent recognition within vehicle cabins. Similarly, \cite{royDoScenesAutonomousDriving2024} introduced the doScenes dataset, pairing 1{,}000 nuScenes driving clips with natural-language instructions to enable voice-driven AV planning. However, neither of these datasets specifically addresses unconstrained conversational exchanges between users and SAV agents, such as analyzing how users perceive the sentiment or appropriateness of SAV responses to their requests.

Meanwhile, large language models (LLMs) like ChatGPT, have emerged as powerful tools capable of generating human-like conversational experiences across diverse domains, including healthcare, education, and intelligent vehicles \cite{thirunavukarasuLargeLanguageModels2023, huaUseLargeLanguage2024, duChatChatGPTIntelligent2023, yuksekgonulOptimizingGenerativeAI2025}. Using LLMs for both conversational data collection and analysis enables granular measurements of user sentiment like polarity and subjectivity directly from dialogue \cite{krugmannSentimentAnalysisAge2024}.

In this study, we present a novel dataset comprising conversational interactions between 50 participants and four different SAE Level 5 SAV agents, simulated using GPT-3.5 turbo with varying prompting strategies. Our dataset includes:

\begin{itemize} 
    \item \textbf{Conversational Textual Data}: 2{,}136 request-response exchanges, accompanied by open-ended and interview insights. 
    
    \item \textbf{Empirical Survey Data}: Structured post-interaction responses capturing user perceptions of a range of psychological factors, and self-reported sentiment measures (Polarity and Subjectivity) toward SAV responses. 
\end{itemize}

To illustrate the utility of our dataset, we provide two benchmark case studies:

\begin{enumerate} 
    \item \textbf{Case Study 1 — SAV Acceptance Analysis} (Empirical Survey Data): We apply the Machine‑Learning‑Chord‑Diagram framework proposed by \cite{guoNewFrameworkPredict2025}, employing random forest modeling to analyze and visualize the most influential item-level predictors of service quality and intention to use SAVs.

    \item \textbf{Case Study 2 — Sentiment Analysis} (Conversational Textual Data): We use an LLM-based sentiment classifier on individual conversational exchanges, extracting metrics (minimum, maximum, mean, median) for polarity and subjectivity. We then benchmark performance against a traditional lexicon-based sentiment analysis tool (TextBlob) and self-reported sentiment scores. 
\end{enumerate}

The remainder of this paper is organized as follows. Section II details our dataset creation process and methodology. Section III presents the first case study, analyzing acceptance predictors derived from survey data. Section IV evaluates LLM-based sentiment analysis performance in the second case study. Finally, Section V concludes with a discussion of study limitations and recommendations for future research.


\section{Dataset \& Methodology}

This section describes the design of the user study, the data collection approach, the structure of the resulting dataset, and details regarding data accessibility.

\subsection{User Study Design and Data Collection}

The user study was developed to investigate how different conversational styles and prompting strategies influence user perceptions, specifically regarding Psychological Ownership (PO), Anthropomorphism (A), Quality of Service (QoS), Disclosure Tendency (DT), Perceived Enjoyment (PE), Behavioral Intention (BI), and the sentiment (Polarity and Subjectivity) associated with SAV responses. Each participant interacted with four simulated SAE Level 5 SAV agents, powered by OpenAI’s \texttt{gpt-3.5-turbo} model. The SAV user interface (UI) was built in Python using the \texttt{gradio} package and integrated with the OpenAI API \cite{gradio, openaiOpenAIPythonAPI2024}. The SAV agents were differentiated by distinct prompting strategies as follows:

\begin{itemize} 
    \item \textbf{SAV 1 (Standard/Control):} Provided baseline functionality such as navigation assistance, climate control adjustments, and media management, without additional personalization.
    
    \item \textbf{SAV 2 (Standard + Psychological Ownership):} Enhanced interactions designed explicitly to foster users’ sense of ownership over the vehicle.

    \item \textbf{SAV 3 (Standard + Anthropomorphism):} Participants selected a preferred anthropomorphic personality (e.g., friendly, sassy, or cool), facilitating personalized and human-like interactions.

    \item \textbf{SAV 4 (Combined PO + Anthropomorphism):} Integrated both psychological ownership and anthropomorphic strategies, aiming to reinforce personalization and ownership engagement simultaneously.
\end{itemize}

All SAV agents were explicitly instructed to comply with Australian national traffic laws and specific regulations applicable within the State of Victoria, aiming for realistic and safe responses throughout the interactions. The exact wording of the prompts is provided in the dataset documentation.

Data collection was conducted with ethical approval from the Monash University Human Research Ethics Committee (MUHREC)\footnote{Project ID: 40485}. The user studies were held from April 30, 2024, to July 2, 2024. A total of 50 participants were recruited from an Australian university community and external networks. 
The study employed a within-subjects experimental design, meaning each participant interacted with all four SAV agents. To minimize potential ordering effects or bias due to interaction sequence, the order in which participants experienced each SAV agent was randomized.

Descriptive statistics, theoretical background and literature on the psychological factors investigated, a detailed overview of the prompt design, statistical comparisons of the effects of psychological ownership and anthropomorphic strategies on user experience, and qualitative analysis of participants’ feedback on perceived psychological ownership are provided in \cite{guoExploringHumanSAVInteraction2025}. The findings indicated that SAV4, which combined both strategies, was perceived as more human-like and elicited more positive user responses overall. In this paper, we present the complete study design, release the full de-identified conversational dataset, and analyze the role of sentiment in human–SAV interactions through two benchmark case studies.

\begin{figure}[htbp]
    \centering
    \includegraphics[width=\columnwidth]{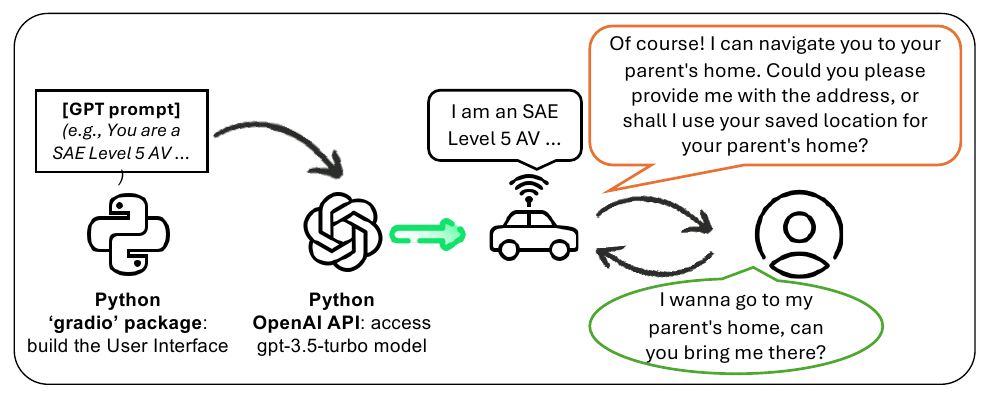}
    \caption{System architecture of the simulated SAV agent.}
    \label{fig:system}
\end{figure}

The system architecture for conversational interactions is illustrated in Fig.~\ref{fig:system}.

Prior to their interactions, participants completed a pre-interaction survey capturing their prior experiences with AI systems and their preferred anthropomorphic personality style for the SAV (Cool \& sophisticated; Engaging \& friendly; or Sassy \& tired). Preferences expressed in this survey informed the configuration of SAV3 and SAV4 during interactions.

A set of 15 in-vehicle interaction commands adapted from those commonly used in existing vehicles such as the Tesla Model Y \cite{noauthor_model_2023} was provided to participants as example requests. These commands covered key vehicle functions, including climate control, vehicle settings, navigation, media controls, and contact management. A complete list of these commands is available in the dataset documentation. Participants were also encouraged to interact with the SAVs freely and make requests in their own words, as they would naturally do in real-life scenarios. 

After each SAV interaction, participants completed a structured post-interaction survey designed to measure perceptions across the selected psychological factors. Additionally, participants were briefly interviewed to capture qualitative feedback on their experiences, particularly perceptions of psychological ownership relative to previous SAV interactions. All conversational exchanges were recorded and stored.

At the end of the entire interaction sequence (all four SAVs), participants completed a final survey collecting demographic information and responded to open-ended questions exploring features that contribute to psychological ownership, perceptions about sharing personal information and other additional insights or feedback.

The complete survey instrument used in this study can be found in the released dataset folder.

Overall, the collected data comprised: 
\begin{itemize} 
    \item 2{,}136 conversational exchanges (request–response pairs) between users and SAV agents;
    \item 200 structured post-interaction survey responses (four responses per participant);
    \item 50 sets of open-ended qualitative statements or interview transcripts (one per participant). 
\end{itemize}

Collected data was anonymized to protect participant privacy and analyzed inline with the approved ethics protocol. Even after broad recoding, several demographic records remained unique (k = 1), and so the dataset failed k-anonymity and we removed the demographic section from the public release \cite{sweeneyACHIEVINGKANONYMITYPRIVACY2002}. The demographic overview is available in \cite{guoExploringHumanSAVInteraction2025}.

\subsection{Data Structure and Availability}

The fully de-identified dataset and accompanying documentation will be publicly accessible via the Monash Bridges repository, released under a CC BY 4.0 license. All files are provided in widely used accessible formats (.xlsx, .pdf, .pptx), facilitating easy reuse and analysis by the research community.

\subsubsection{Data Structure}

The dataset comprises the following:

\begin{itemize} 
    \item \textbf{Data\_human-SAV\_interaction.xlsx:} Contains 2{,}136 conversational exchanges (request–response pairs) between users and SAV agents.

    \item \textbf{Data\_Survey1-4.xlsx:} Structured responses from post-interaction surveys (four per participant), capturing psychological ownership, anthropomorphism, quality of service, disclosure tendency, perceived enjoyment, behavioral intention, and self-reported sentiment measures (polarity and subjectivity) regarding SAV responses.

    \item \textbf{Data\_survey\_open\&end.xlsx:} Pre-interaction survey data assessing participants' prior experience with AI systems, their preferred anthropomorphic SAV personalities (e.g., friendly, cool), and responses to the final survey completed at the end of the study.

    \item \textbf{Data\_interview\_and\_openendedQ.xlsx:} Transcribed qualitative data from participant interviews and responses to open-ended survey questions, capturing nuanced user feedback and perceptions.

    \item \textbf{Prompts.pptx:} Prompts used for developing each of the four SAV agents.

    \item \textbf{Sample\_user\_input.pdf:} Representative examples of user commands, illustrating navigation requests, comfort adjustments, and emotional prompts.

    \item \textbf{Survey.pdf:} Full survey instrument utilized during the study.
\end{itemize}

\subsubsection{Access and Reuse}

The dataset will be hosted for open and persistent access.

\begin{itemize} 
    \item \textbf{URL:} \url{https://doi.org/10.26180/29486447}
    \item \textbf{DOI:} 10.26180/29486447 
    \item \textbf{License:} CC BY 4.0 (Creative Commons Attribution) 
\end{itemize}

\section{Case study 1: Item importance in predicting SAV acceptance}

\subsection{Motivation}

Identifying specific aspects of user experience that most significantly influence SAV acceptance is crucial for designing interfaces and interaction strategies that foster user satisfaction and sustained use. Prior research has examined various psychological factors influencing public acceptance of SAVs and human-computer interaction more broadly, including Psychological Ownership (PO) \cite{leeAutonomousVehiclesCan2019, dai_impacts_2021}, Anthropomorphism (A) \cite{wuDeepSuperficialAnthropomorphism2023}, Disclosure Tendency (DT) \cite{chengGoodBadUgly2022}, Quality of Service (QoS) \cite{dai_impacts_2021}, Perceived Enjoyment (PE) \cite{songPredictorsConsumersWillingness2021}, and sentiment (Polarity and Subjectivity) \cite{wankhadeSurveySentimentAnalysis2022}.

Most previous studies have applied traditional statistical methods, such as Structural Equation Modeling (SEM), relying primarily on aggregate scores to explore these relationships. Recent research, however, has demonstrated the effectiveness of machine learning methods, such as Neural Networks and Random Forests, in predicting user acceptance of AVs \cite{lirui-094, guoNewFrameworkPredict2025}. To overcome limitations associated with traditional SEM, particularly its inability to identify item-level predictors, \cite{guoNewFrameworkPredict2025} introduced a predictive modeling framework combined with chord diagrams, enabling researchers to visualize the relative importance of specific questionnaire items. This approach uncovers fine-grained insights into drivers of various psychological factors and behavioral intentions, providing actionable information for SAV interface designers and service providers.

\subsection{Prediction and visualization framework}

\begin{figure}
    \centering
    \includegraphics[width=0.8\linewidth]{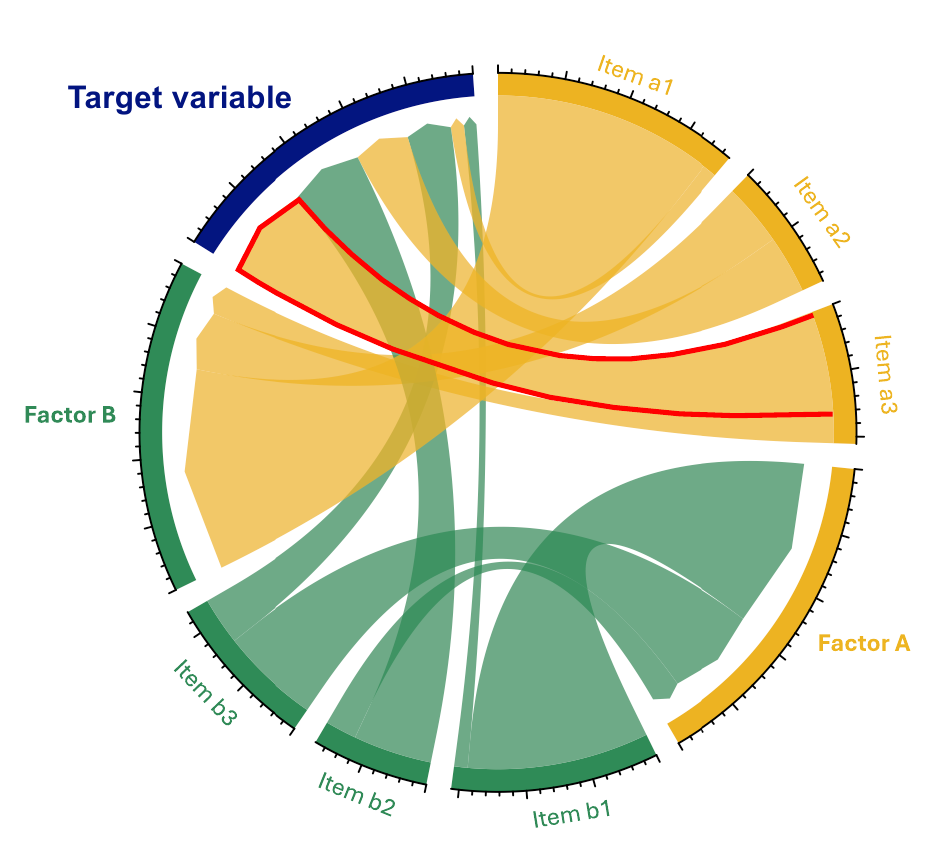}
    \caption{Example chord diagram illustrating item-level importance in predicting a target variable. Two latent factors are shown, each comprising multiple items used as predictors. The width of each arc represents the relative importance of the item, normalized such that the total importance across all predictors sums to 100\%. In this example, Item A3 emerges as the most influential predictor among the six items.}
    \label{fig:cir_example}
    \vspace{-4mm}
\end{figure}

In this study, we applied the predictive modeling and visualization framework proposed by \cite{guoNewFrameworkPredict2025} to our post-interaction survey dataset. The framework predicts each psychological factor by averaging its corresponding survey items using Random Forest models. For example, the Quality of Service factor is computed from items QoS1–QoS3, which measure user's perceived SAV service quality (QoS1), communication pleasantness (QoS2), and recommendation likelihood (QoS3). Full item descriptions are provided in the dataset documentation. The relative importance of each item in predicting these factor-level scores was visualized through chord diagrams. An example of how to interpret these diagrams is given in Fig.~\ref{fig:cir_example}.

\begin{figure*}[!t]
    \centering
    
    \begin{subfigure}[b]{0.48\textwidth}
        \centering
        \includegraphics[width=\linewidth]{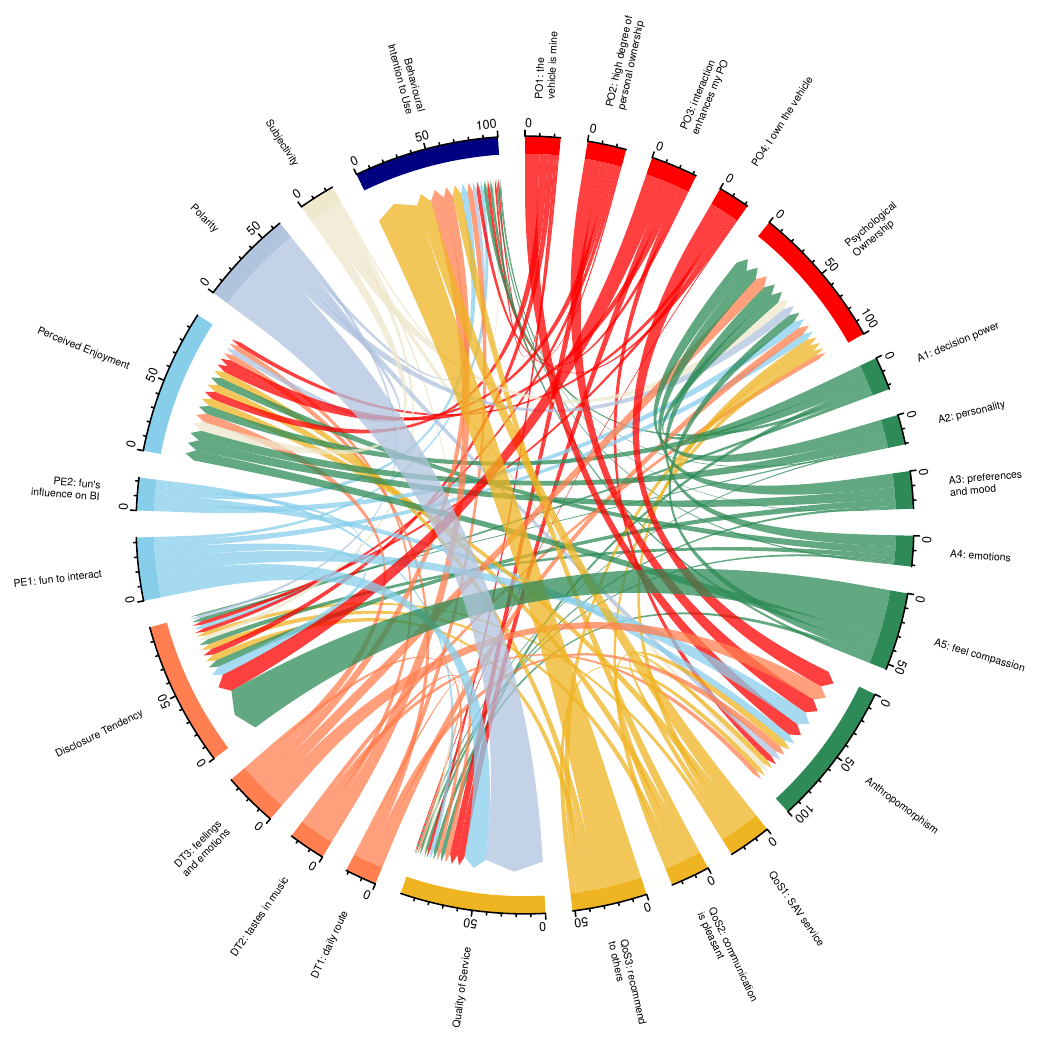}
        \caption{SAV1 - Standard.}
        \label{fig:cir_SAV1}
    \end{subfigure}
    \hfill
    \begin{subfigure}[b]{0.48\textwidth}
        \centering
        \includegraphics[width=\linewidth]{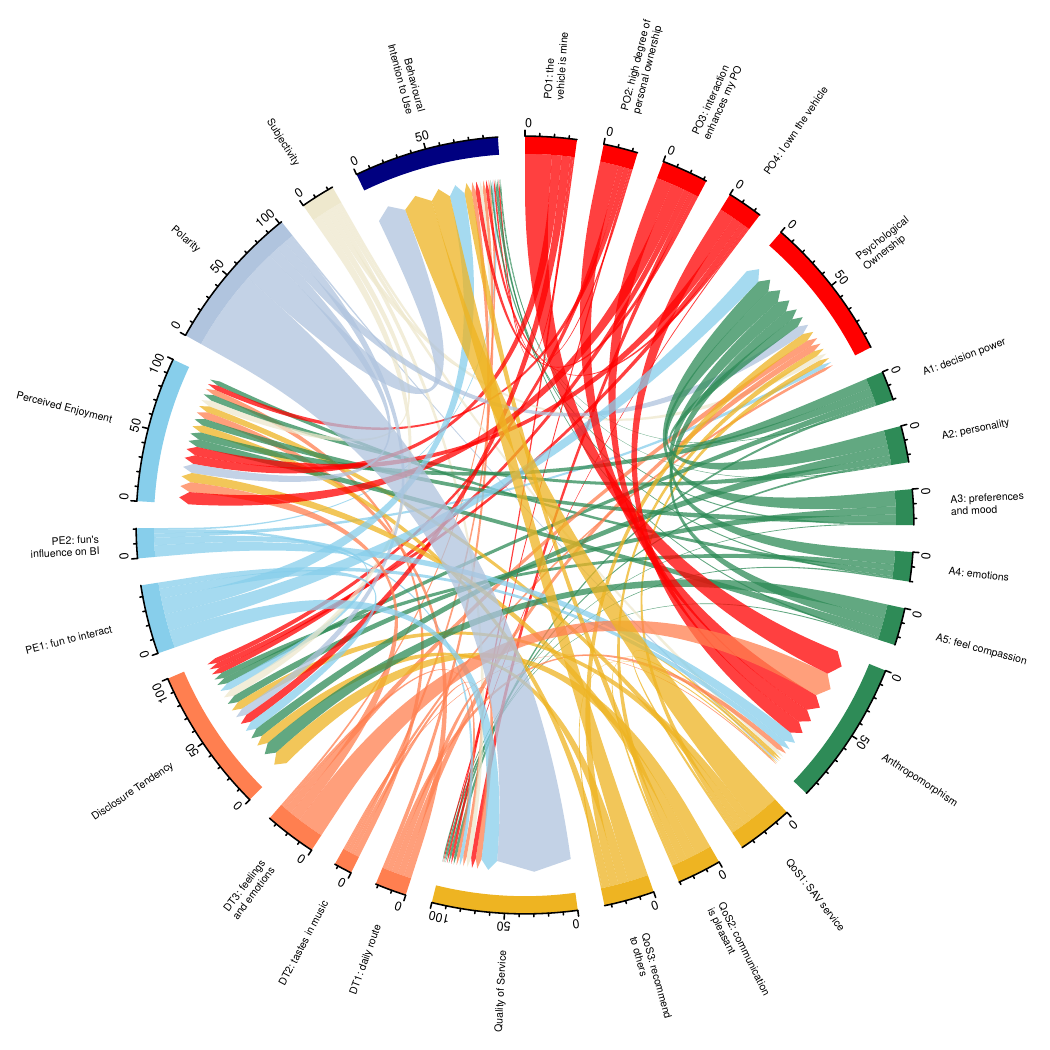}
        \caption{SAV2 - Standard + Psychological Ownership.}
        \label{fig:cir_SAV2}
    \end{subfigure}
    
    
    \begin{subfigure}[b]{0.48\textwidth}
        \centering
        \includegraphics[width=\linewidth]{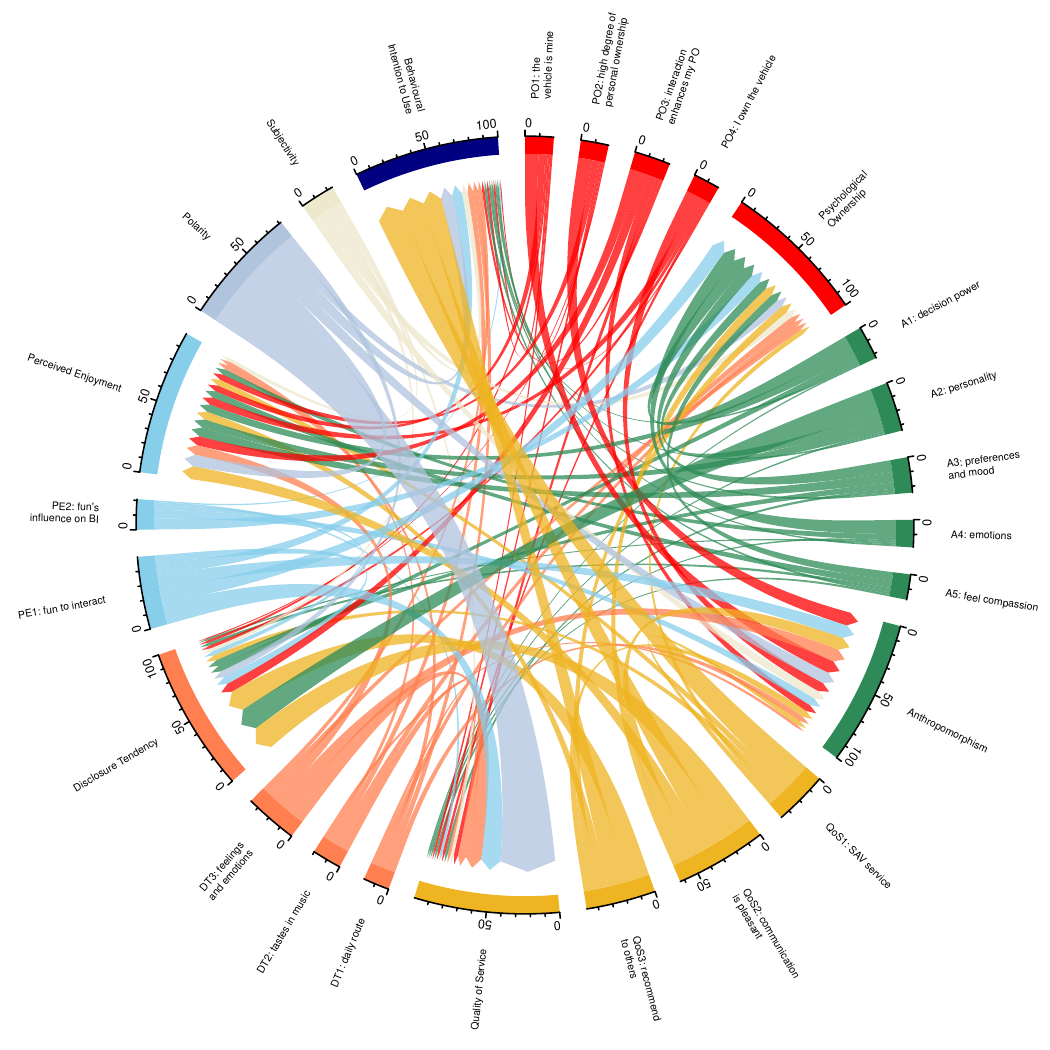}
        \caption{SAV3 - Standard + Anthropomorphism.}
        \label{fig:cir_SAV3}
    \end{subfigure}
    \hfill
    \begin{subfigure}[b]{0.48\textwidth}
        \centering
        \includegraphics[width=\linewidth]{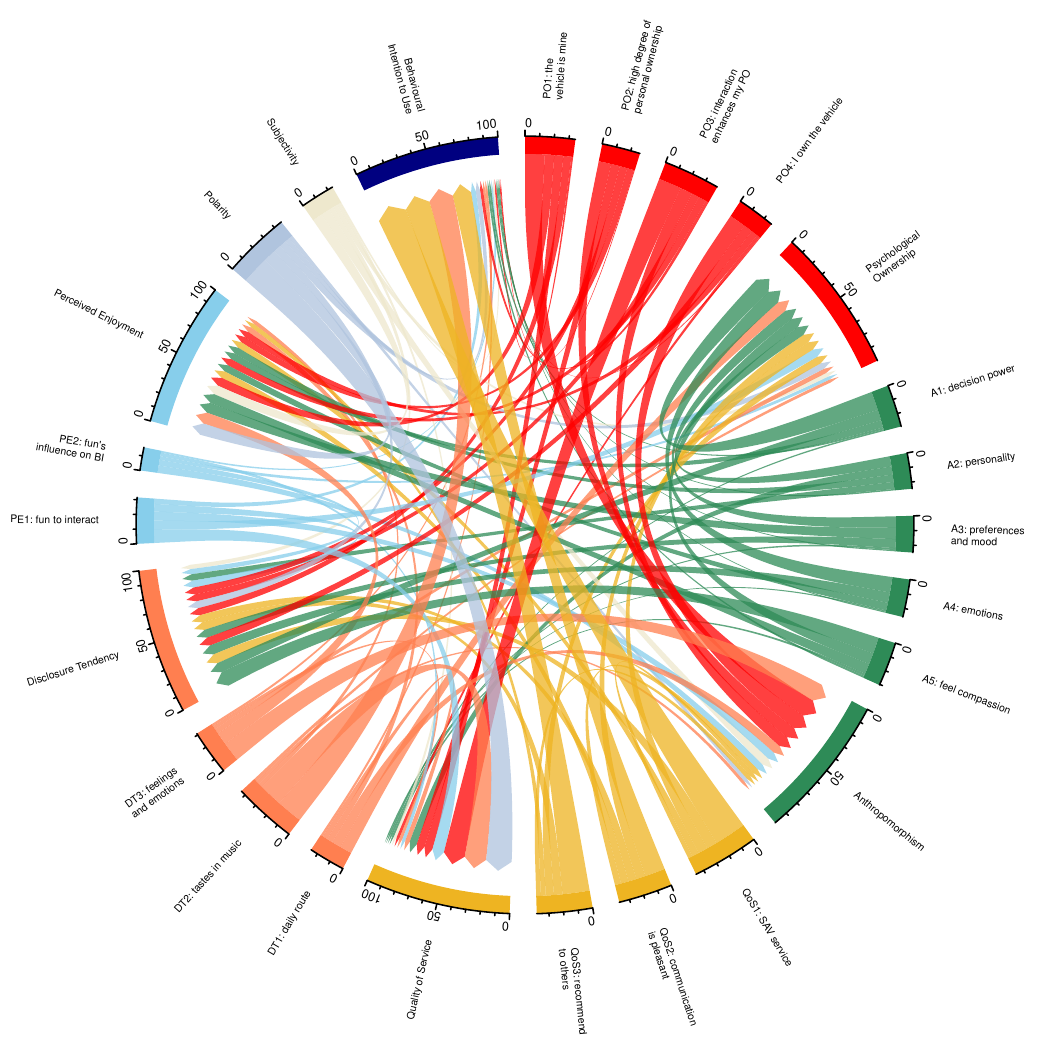}
        \caption{SAV4 - Standard + PO + A}
        \label{fig:cir_SAV4}
    \end{subfigure}
    
    \caption{Chord Diagrams Showing Relative Importance of Predicting Target Factors. The arrows (chords) connect the predictor items (start nodes) and the target variables (end nodes). The width of the chords indicates the relative importance of each predictor item. The sum of the relative importance of each target variable (i.e., the overall questions) is 100\%.}
    \label{fig:cir_plot}
    \vspace{-5mm}
\end{figure*}

Comparisons across the four SAVs reveal variations in user perceptions attributable to different prompting strategies. The resulting relative item importance is shown visually in Fig.~\ref{fig:cir_plot}.

\subsection{Psychological ownership prompt effect - SAV1 vs. SAV2}

When predicting Behavioral Intention to use SAVs, the relative importance of the factor items for SAV1 and SAV2 is visualized in Fig.~\ref{fig:cir_SAV1} and~\ref{fig:cir_SAV2}, respectively. In SAV1, the most influential item is QoS3 (recommend to others, with a relative importance of 32.2\%), followed by QoS1 (SAV service, 11.5\%) and DT2 (tastes in music, 10.5\%). In contrast, SAV2 demonstrates a different pattern, with Polarity emerging as the most influential item (22.4\%), followed by QoS1 (SAV service, 22.0 \%) and QoS3 (recommend to others, 14.7\%). These findings suggest that integrating psychological ownership prompts increases the influence of the Polarity of SAV responses, indicating that users pay more attention to the sentiment of the SAV responses. In both groups, Polarity plays a key role in predicting Quality of Service (QoS), with relative importance of 44.7\% and 57.9\% respectively, suggesting that users’ perceptions of SAV responses significantly shape their overall evaluation of the service.

Similarly, when predicting user's Disclosure Tendency, A5 (feel compassion, 34.4\%) emerged as the most important predictor for SAV1, followed by PO3 (interaction enhances my PO, 14.2\%). While for SAV2, QoS2 (communication is pleasant, 11.4\%) and A5 (feel compassion, 10.1\%) are the most influential items. This reveals that after integrating psychological ownership prompts, although the beliefs of SAV can feel compassion is still important, the pleasantness of communication becomes a more significant factor.

Another notable difference arises when predicting Psychological Ownership. In SAV1, the most influential items are A2 (personality, 10.0\%), A3 (preferences and mood, 9.9\%), and DT3 (feelings and emotions, 9.1\%). In contrast, for SAV2, the most influential items are PE1 (fun to interact, 12.2\%), A3 (preferences and mood, 11.1\%), and A2 (personality, 11.0\%). These results indicate that while anthropomorphism-related items, particularly beliefs about the SAV having its own personality and preferences, are important in predicting psychological ownership for both SAV groups, other factors differ. In SAV1, the willingness to disclose personal feelings and emotions (DT3) plays a more important role, whereas in SAV2, the enjoyment of interaction (PE1) becomes a more critical factor.

\subsection{Anthropomorphism prompt effect - SAV1 vs. SAV3}

When comparing the relative importance of factor items for SAV1 and SAV3 (Fig.~\ref{fig:cir_SAV3}), diverse patterns emerge. In predicting Behavioral Intention to use SAVs, the most influential items for SAV3 are QoS1 (SAV service, 20.5\%), QoS2 (communication is pleasant, 17.3\%), and QoS3 (recommend to others, 14.1\%). Compared to SAV1, participants appear to place greater emphasis on whether communication with the SAV is pleasant. Similar to the psychological ownership prompt effect, the Polarity of SAV responses remains the most important predictor of QoS in both groups, with a relative importance of 44.7\% in SAV1 and 43.0\% in SAV3.

In predicting Disclosure Tendency, the most influential items in SAV3 are QoS2 (communication is pleasant, 19.0\%), A2 (personality, 17.7\%), and QoS3 (recommend to others, 16.6\%), which differ from the most important predictors in SAV1 (A5, PO3, PE1). This shift suggests that, after integrating anthropomorphism prompts, the QoS items, particularly the pleasantness of communication and the likelihood of recommending the SAV to others, become more critical in predicting users’ willingness to disclose personal information and emotions.

Surprisingly, when predicting Psychological Ownership, the most influential items in SAV3 are similar to but distinct from those in SAV2. For SAV3, the most important predictors are PE1 (fun to interact, 10.7\%), A4 (emotions, 10.2\%), and A1 (decision power, 9.5\%). Compared to the control group (i.e., SAV1, where A2, A3, and DT3 are the most important predictors), these results indicate that, following the integration of anthropomorphism prompts, factors such as the enjoyment of interaction and beliefs about the SAV having its own emotions and decision-making power play a more significant role in predicting Psychological Ownership.

\subsection{Combined prompt effect - SAV4}

The combined group, SAV4 (Fig.~\ref{fig:cir_SAV4}), integrates both PO and A prompts. When predicting Behavioral Intention to use SAVs, the most influential items are QoS1 (SAV service, 23.0\%), QoS3 (recommend to others, 19.5\%), and DT2 (tastes in music, 18.9\%). These results are similar to SAV1 but show a higher relative importance for QoS1 and DT2, and a lower relative importance for QoS3. This suggests that while the quality of service and the likelihood of recommending the SAV remain important predictors across all four SAVs, the integration of both prompts enhances the influence of users’ willingness to disclose their music preferences, unlike the focus on Polarity seen in SAV2 and SAV3.

When predicting Psychological Ownership, although Anthropomorphism items still play a significant role, the willingness to disclose daily routes (DT1, 8.7\%) emerged as the third most important predictor. Interestingly, when predicting QoS, while the Polarity of SAV responses remains the most influential item (20.1\%), its relative importance is lower than in SAV1, SAV2, and SAV3. This suggests that while the sentiment of SAV responses continues to be a critical factor, the integration of psychological ownership and anthropomorphism prompts shifts importance toward other factors, such as willingness to disclose music preferences (DT2, 16.9\%) and the perception of interaction-enhanced psychological ownership (PO3, 16.6\%).

Notably, the integration of psychological ownership prompts (in SAV2 and SAV4) increases the overall relative importance of the Psychological Ownership factor in predicting other target variables. The relative importance is higher in SAV2 (116.1\%) and SAV4 (126.2\%) compared to SAV1 (105.7\%) and SAV3 (82.4\%). This finding suggests that psychological ownership prompts reinforce the influence of psychological ownership on users’ perceptions of other factors, such as quality of service and anthropomorphism.

\section{Case study 2: using LLMs as sentiment analysis tools}

\subsection{Motivation}

Case Study~1 showed that conversational \textbf{sentiment polarity} is the strongest item-level predictor of perceived Quality of Service (QoS) across all four SAVs, and that QoS emerged as the most critical factor predicting Behavioral Intention (BI) to use SAVs. Even when aggregating importance at the factor-level, polarity remained notably influential, especially for SAV1, SAV2, and SAV3. These results point to a practical next step: \emph{if polarity drives user perceptions, can we measure it automatically?} Specifically, can we reliably derive polarity and subjectivity measures (i.e., sentiment analysis metrics) directly from conversational text to predict user ratings? Addressing this question is the focus of Case Study~2, where we benchmark two sentiment-analysis approaches and evaluate how well their outputs track participants’ self-reported sentiment.

Sentiment analysis integrates natural language processing (NLP), computational linguistics, and text analytics to extract and interpret emotional tone from textual data \cite{bonta_comprehensive_2019}. Traditional sentiment tools, such as TextBlob, adopt a lexicon-based method suitable for sentence-level evaluations. TextBlob offers straightforward interfaces for various NLP tasks -- including sentiment scoring (polarity and subjectivity), part-of-speech tagging, and noun phrase extraction \cite{loria_textblob_2022}. Polarity scores range from –1 (extremely negative) to 1 (extremely positive), while subjectivity ranges from 0 (highly objective) to 1 (highly subjective). With recent advancements in LLMs, researchers have begun exploring their effectiveness in sentiment analysis tasks. For example, \cite{krugmannSentimentAnalysisAge2024} demonstrated that zero-shot LLM approaches, such as GPT-3.5 and GPT-4, can match or exceed traditional transfer learning methods across various sentiment benchmark datasets.

In this case study, we evaluate two sentiment detection strategies on conversational data collected during human–SAV interactions: a traditional lexicon-based tool (TextBlob) and an LLM-based sentiment analysis method. We subsequently assess the alignment between these automated sentiment scores and the sentiment perceptions participants reported in post-interaction surveys.

\subsection{Experiment design and evaluation metrics}

We used the \texttt{TextBlob} package (version 0.19.0) in Python. For the LLM-based sentiment analysis, we employed \texttt{gpt-4o-mini} using a zero-shot prompt, as shown below.

Both sentiment analysis methods were applied independently to every SAV response within each participant's conversation. For instance, if Participant P01 completed 15 request-response exchanges with SAV1, both methods produced 15 sentiment evaluations (each containing polarity and subjectivity scores). Aggregate statistics (minimum, maximum, mean, median, and mode) were then computed from these evaluations to create representative sentiment measures for each participant–SAV interaction.

\begin{tcolorbox}[promptbox, title=Zero‑Shot LLM Sentiment‑Analysis Prompt]
You are an advanced Sentiment Analysis Model for evaluating responses from a Shared Autonomous Vehicle (SAV). For each SAV response, analyze its explicit and implied sentiment, and then output two scores:
\begin{enumerate}
  \item Polarity Score: A number from \(-1\) (extremely negative) to \(1\) (extremely positive).
  \item Subjectivity Score: A number from \(0\) (highly objective) to \(1\) (highly subjective).
\end{enumerate}
Instructions:\\
Evaluate its overall emotional tone and degree of personal bias.\\
Return only a valid JSON object with the keys "Polarity Score" and "Subjectivity Score” - no additional text.\\
Example JSON structure: {"Polarity Score": <number>, "Subjectivity Score": <number> }.
\end{tcolorbox}

\begin{figure}[htbp]
  \centering

  \begin{subfigure}[t]{\columnwidth}
    \centering
    \includegraphics[width=0.9\columnwidth]{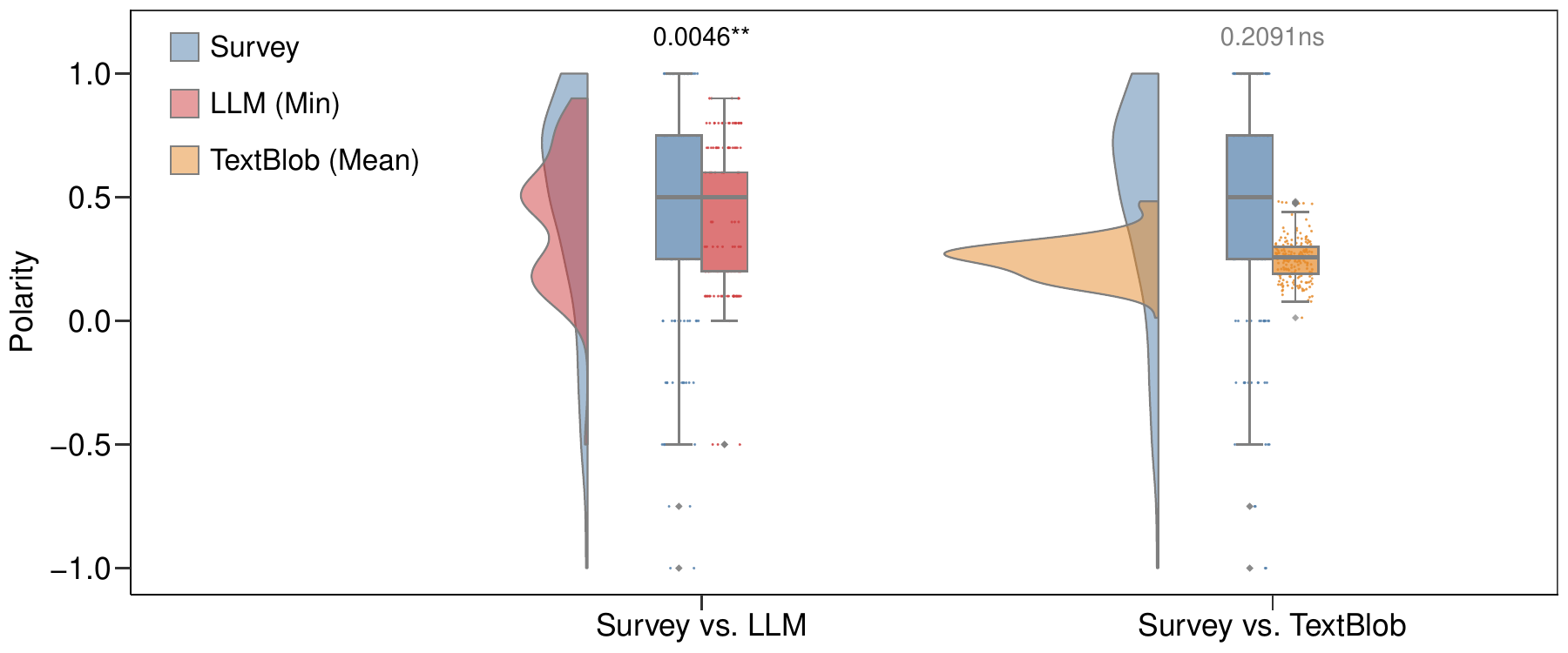}
    \caption{Polarity.}
    \label{fig:polarity_comparison}
  \end{subfigure}

  \vspace{0.5em} 

  \begin{subfigure}[t]{\columnwidth}
    \centering
    \includegraphics[width=0.9\columnwidth]{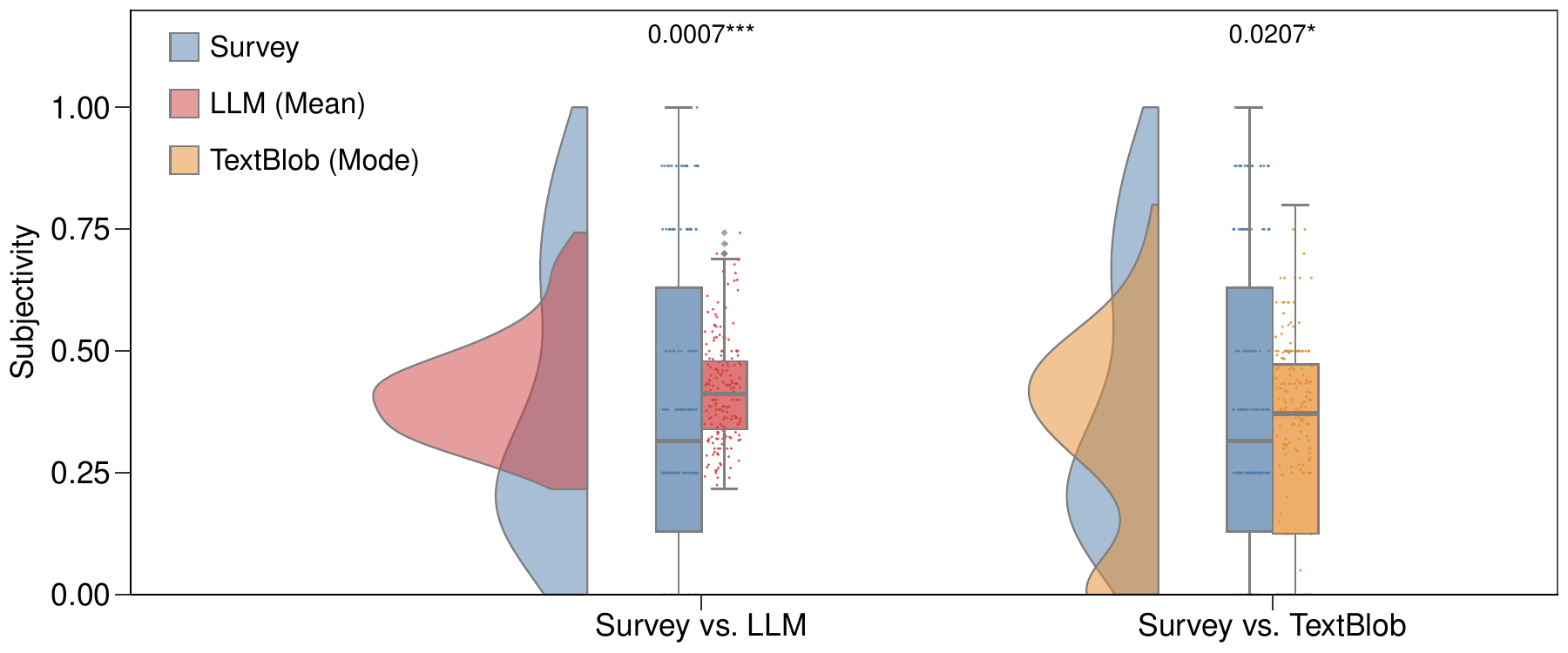}
    \caption{Subjectivity.}
    \label{fig:subjectivity_comparison}
  \end{subfigure}

  \caption{%
    Raincloud plots comparing survey scores to the highest-correlating sentiment features (among min, max, mean, median, mode) from LLM and TextBlob. Each combines density, boxplot, and jittered points, with Spearman p-values annotated. Kernel density estimates were computed using Scott’s rule for bandwidth selection.
  }
  \label{fig:raincloud_comparison}
  \vspace{-3mm}
\end{figure}

The non-parametric Spearman’s rank correlation coefficient was employed to assess the relationship between the aggregated sentiment metrics and the corresponding survey-based sentiment ratings provided by participants \cite{SpearmansRankOrderCorrelation}. Correlation outcomes are summarized in Table~\ref{tab:spearman_merged}. The sentiment features with the highest correlation from each method were visualized against survey-based distributions in density plots (Figs.~\ref{fig:polarity_comparison} and~\ref{fig:subjectivity_comparison}).

\begin{table}[htbp]
  \centering
  \caption{Spearman Correlation (r) Between Computed Sentiment Features and Survey-Based Ratings. Significance levels: *** $p<0.001$, ** $p<0.01$, * $p<0.05$.}
  \label{tab:spearman_merged}
  \footnotesize
  \setlength{\tabcolsep}{8pt}
  \begin{tabular}{@{}lcc@{}}
    \toprule
    \textbf{Method} 
      & \textbf{$r_{\mathrm{polarity}}$} 
      & \textbf{$r_{\mathrm{subjectivity}}$} \\
    \midrule
    \texttt{llm\_min}        & \textbf{0.199}**  & 0.010 \\
    \texttt{llm\_mean}       & \textbf{0.182}*   & \textbf{0.237}** \\
    \texttt{textblob\_mean}  & 0.089    & 0.032  \\
    \texttt{textblob\_min}   & 0.089    & 0.006  \\
    \texttt{llm\_median}     & 0.075    & \textbf{0.217}**   \\
    \texttt{textblob\_median}& 0.060    & -0.056   \\
    \texttt{textblob\_max}   & 0.021    & 0.076   \\
    \texttt{textblob\_mode}  & 0.002    & \textbf{0.163}*   \\
    \texttt{llm\_mode}       & -0.001   & 0.131  \\
    \texttt{llm\_max}        & -0.089   & \textbf{0.145}*  \\
    \bottomrule
  \end{tabular}
\end{table}
\vspace{-3mm}

\subsection{Findings and Implications}

Our results indicate that the LLM-based minimum polarity measure (\texttt{llm\_min}) achieved the strongest correlation with users’ self-reported polarity (\(r = 0.199\), \(p = 0.005\)) (Table~\ref{tab:spearman_merged}). This result is visually reinforced by the density plot in Fig.~\ref{fig:polarity_comparison}, which shows that the LLM polarity distribution more closely matches the survey‑based distribution than TextBlob’s distribution. Similarly, for subjectivity, the LLM-derived mean score (\texttt{llm\_mean}) had the highest correlation with survey-based subjectivity ratings (\(r = 0.237\), \(p = 0.001\)) (Table~\ref{tab:spearman_merged}), outperforming TextBlob’s best-performing measure (\texttt{textblob\_mode}, \(r = 0.163\), \(p = 0.021\)). In both sentiment dimensions, the LLM-based method consistently outperformed TextBlob at statistically significant levels \(p<0.05\).

Despite modest correlation magnitudes (\(r \approx 0.2\)),  these results offer several meaningful insights:
\begin{enumerate}
  \item \textbf{Holistic user evaluation}: Users appear to rate sentiment based on the overall conversational experience, encompassing factors beyond individual responses. Modest correlation values imply that text-only sentiment metrics alone are insufficient for capturing the complete user experience, which likely includes context, multimodal cues, and interpersonal interaction nuances.
  \item \textbf{Sensitivity to extreme sentiment}: The prominence of the minimum polarity score indicates users are likely affected by the most negative (``worst-case'') exchanges during the interactions.
  \item \textbf{LLM efficacy}: Even with a simple zero‑shot prompt, the LLM sentiment method significantly outperformed TextBlob, demonstrating strong potential for conversational sentiment analysis.
\end{enumerate}

These findings underscore the potential of LLM-based sentiment analysis as a powerful and efficient tool for evaluating conversational interactions. However, the modest correlation between LLM-derived sentiment scores and participants’ self-reported ratings also reveals its current limitations. Surprisingly, the LLM-based tool did not accurately reflect users’ perceived sentiment toward SAV responses. Two possible explanations emerge: First, as previously discussed, participants' self-ratings may have reflected their aggregate experience over the entire interaction, particularly influenced by notably negative exchanges such as the refusal of a request, rather than the sentiment of isolated SAV responses. Participants may disproportionately emphasize negative experiences or interactions that fail their expectations, a phenomenon in social psychology whereby negative experiences have a stronger impact on overall evaluations than equivalent positive ones \cite{rozinNegativityBiasNegativity2001, xuTwoSidesHeuristics2025}. Second, LLMs may still lack the capacity to fully capture the nuanced sentiment embedded in complex, real-world dialogues. These limitations point to promising directions for future research, including the integration of richer contextual and multimodal cues, such as dialogue history, speech prosody, and facial expressions -- to improve the accuracy and depth of sentiment modeling in human–SAV interactions.

The identified importance of sentiment polarity offers actionable insights for SAV interface designers. As demonstrated in Case Study 1, SAV response polarity plays an important role in influencing perceived service quality. This suggests that LLM-powered sentiment analysis tools could be effectively integrated into autonomous vehicle systems to enable real-time sentiment monitoring. When signs of user dissatisfaction or confusion are detected, the system could proactively respond to negative sentiment dips by deploying adaptive strategies to maintain a positive and stable user experience. These targeted design strategies, grounded in sentiment polarity insights, offer a promising pathway toward emotionally responsive and user-centered SAV interfaces.




\section{Conclusions}

This paper introduced an open-source human–SAV interactions dataset. The dataset features diverse prompting strategies designed to elicit psychological ownership and anthropomorphic engagement, and includes both structured survey responses and rich conversational textual data. It offers insights into key user perceptions such as psychological ownership, anthropomorphism, quality of service, disclosure tendency, perceived enjoyment, behavioral intention, and sentiment.

We demonstrated the utility of this dataset through two benchmark case studies. In Case Study 1, we applied Random Forest modeling to the survey data to examine item-level predictors of SAV acceptance, revealing how different conversational strategies influence user experience. In Case Study 2, we analyzed the conversational data using both a LLM-based sentiment analysis tool and TextBlob, comparing their alignment with user-reported sentiment ratings.


Several promising directions emerge for future work. First, this screen-based prototype could be replicated in realistic driving environments to capture richer interaction cues. Second, future studies should recruit more diverse participants to improve generalization. Third, benchmarking additional LLM architectures and advanced sentiment analysis methods could address potential model biases and improve prediction accuracy. Lastly, randomizing or removing example commands could encourage more realistic and spontaneous user interactions. These extensions will further clarify sentiment’s role in shaping user experience to inform  SAV design.

We invite researchers to further explore this dataset in the context of human–SAV interaction. Future studies could, for example, investigate the predictive power of conversational text in estimating perceived quality of service, or develop more advanced prompting techniques to enhance the performance of LLM-based sentiment analysis, building upon the findings of Case Study 2.


\bibliography{reference}

\end{document}